\documentclass[11pt]{article}
\usepackage[a4paper,margin=1in]{geometry}
\usepackage{amsmath,amssymb,amsthm,amsfonts}
\usepackage{bm}
\usepackage{authblk}
\usepackage{hyperref}
\usepackage{setspace}
\usepackage{graphicx} 
\usepackage{subcaption}
\usepackage{cite}

\usepackage{xcolor}

\title{Resonance-Enhanced Time Reflection at Photonic Temporal Interfaces}
\date{}

\author[1,$\dagger$]{Zeyuan Li}
\author[2,3,$\dagger$]{Hammam Bahurmuz}
\author[2]{Mohamed H. Mostafa}
\author[4]{Mohammad S. Mirmoosa}
\author[5]{Puneet Garg}
\author[5,6,7]{Carsten Rockstuhl}
\author[1]{Xuchen Wang\thanks{Corresponding author: \href{mailto:xuchen.wang@hrbeu.edu.cn}
  {xuchen.wang@hrbeu.edu.cn}}}
\author[2]{Viktar Asadchy}

\affil[1]{College of Physics and Optoelectronic Engineering, Harbin Engineering University, China}
\affil[2]{Department of Electronics and Nanoengineering, Aalto University, Espoo, Finland}
\affil[3]{TUM School of Computation, Information and Technology, Technical University of Munich, Munich, Germany}
\affil[4]{Department of Physics and Mathematics, University of Eastern Finland, P.O.~Box~111, FI-80101 Joensuu, Finland}
\affil[5]{Institute of Theoretical Solid State Physics, Karlsruhe Institute of Technology, Karlsruhe, Germany}
\affil[6]{Institute of Nanotechnology, Karlsruhe Institute of Technology, Karlsruhe, Germany}
\affil[7]{Center for Integrated Quantum Science and Technology, Karlsruhe Institute of Technology, Wolfgang-Gaede-Straße 1, 76131 Karlsruhe, Germany}

\begin{document}
\maketitle

\begingroup
\renewcommand{\thefootnote}{}
\footnotetext{\textsuperscript{$\dagger$}These authors contributed equally to this work.}
\endgroup

\begin{abstract} 
Photonic temporal interfaces enable dynamic control of light fields, yet strong time reflection at optical frequencies remains challenging because conventional approaches demand large material refractive-index changes on ultrafast timescales. Here, we introduce a resonance-assisted mechanism that harnesses both polarization energy accumulated in a dispersive medium prior to the temporal interface and strongly increased energy supplied by the modulation system. We show that, under a specific critical condition, rapidly increasing the material resonance frequency yields orders-of-magnitude stronger time-reflected power flux density than what conventional plasma-frequency modulation provides. To implement this mechanism in optical systems, we identify two routes based on dielectric and plasmonic structural resonances. For the plasmonic route, we develop an analytical effective-medium model of conducting-oxide cylinder arrays, in which localized surface-plasmon resonances transform the constituent Drude response into a geometrically tunable effective Lorentz response. Using cadmium oxide as a representative material, we predict an enhancement exceeding three orders of magnitude in the summed reflected-mode power coefficient relative to the same homogeneous material under the same modest plasma-frequency modulation, even in the presence of realistic losses. These findings establish spatial resonance engineering as an effective route to strong temporal scattering with reduced demands on intrinsic material tunability.
\end{abstract}

\section{Introduction}
The comprehensive control of light has traditionally relied on static spatial structuring, as exemplified by metamaterials, which enables reflection, refraction, localization, and band engineering while preserving frequency during scattering. Recently, it has become increasingly clear that time can serve as an additional degree of freedom for wave control~\cite{engheta2020metamaterials,galiffi2022photonics,asgari2024theory,yu2026fundamentals,mostafa2024temporal}. If the optical properties of a medium are modulated while a wave is propagating inside it, the wave no longer experiences a static background but a dynamically changing environment. The simplest and most fundamental example of such a process is a temporal interface, formed when the electromagnetic properties of a spatially uniform medium change abruptly in time. Unlike a spatial interface, where the frequency is conserved and the wave vector changes, a temporal interface conserves the wave vector but allows the frequency to change~\cite{morgenthaler2003velocity,xiao2014reflection,ortega2023tutorial}. As a result, an incident wave decomposes at a temporal interface into two scattered waves with shifted frequencies: a time-refracted wave that continues along the initial propagation direction, and a time-reflected wave that propagates backward in space with a conjugated phase~\cite{mendoncca2002time}.

Temporal interfaces enable unusual wave phenomena and provide a wide range of intriguing possibilities, including temporal Fabry--Perot interference~\cite{moussa2023observation}, inverse prism effects~\cite{akbarzadeh2018inverse}, temporal aiming~\cite{pacheco2020temporal}, subwavelength imaging~\cite{simovski2025electromagnetic}, antireflection temporal coatings~\cite{ramaccia2020light,pacheco2020antireflection,liberal2023quantum}, direction-dependent wave manipulation~\cite{mirmoosa2024time}, extreme energy transformations~\cite{li2021temporal}, space--time wave routing~\cite{pacheco2025temporal}, polarization engineering~\cite{mostafa2023spin,mirmoosa2024time}, complete polarization conversion~\cite{xu2021complete}, polarization-dependent analog computing~\cite{rizza2023spin}, wave freezing and thawing~\cite{wang2023controlling,pacheco2025holding}, the transformation of surface waves into free-space radiation~\cite{wang2023controlling,shirokova2019scattering}, generation of persistent nanoscale magnetization~\cite{rawat2026generation}, angular-dependent inhibition of photon production~\cite{vazquez2023shaping}, photon-pair destruction, and vacuum state generation~\cite{mirmoosa2025quantum}. A particularly important extension is the photonic time crystal, in which a periodic modulation of the material properties creates a sequence of temporal interfaces~\cite{galiffi2022photonics,yin2022floquet,sharabi2022spatiotemporal,lustig2023photonic,wang2023metasurface,wang2025expanding}. Just as spatial photonic crystals arise from interference between waves scattered by spatial interfaces, photonic time crystals rely on repeated interference between time-reflected and time-refracted waves generated at successive temporal interfaces. This interference opens momentum band gaps rather than frequency band gaps, enabling nonresonant temporal amplification, waveform shaping, momentum gaps for nonuniform waves, and temporal topological states~\cite{lyubarov2022amplified,hwang2024waveform,dong2025nonuniform,lustig2018topological,xiong2025observation}.

Despite their fundamental importance, time-reflected waves, unlike time-refracted waves, remain difficult to generate with appreciable amplitude because they generally require a large, nonadiabatic material change within a fraction of the wave period. Observations have therefore been largely limited to water waves~\cite{bacot2016time}, cold atoms~\cite{dong2024quantum}, microwave transmission lines~\cite{moussa2023observation,jones2024time}, acoustic and elastic systems~\cite{kim2024temporal,wang2025elastic}, and synthetic photonic dimensions~\cite{feis2025space,long2023time}. Very recently, a backward optical signal was reported as evidence of time reflection in an optically pumped CdO film, where a finite-thickness ENZ layer underwent pump-induced space--time modulation~\cite{segal2026about}. This experiment represents an important step toward probing optical temporal scattering, but it leaves unresolved the central challenge of generating the strong and controllable time-reflected waves required for practical temporal-wave-control applications. More broadly, no systematic framework currently identifies how to engineer realistic dispersive material responses to maximize time reflection at high frequencies.

In this work, we develop a resonance-assisted strategy for maximizing time reflection at a single temporal interface. Combining analytical theory with full-wave finite-difference time-domain (FDTD) simulations, we demonstrate that resonance-frequency modulation can produce substantially stronger time reflection than plasma-frequency modulation for comparable modulation depths. This enhancement originates from both the initially stored polarization energy and the work supplied by the modulation mechanism, which can become very large near the initial material resonance. A mechanical oscillator analogy provides an intuitive explanation of this energy transfer. Independent analyses of the instantaneous energy balance and the time-averaged post-switch energy redistribution yield consistent results, supported by the FDTD simulations. 
To translate these findings into realistic platforms, we propose two implementation routes based on dielectric and plasmonic structural resonances. For the latter, we develop an analytically tractable Maxwell--Garnett effective-medium model of a two-dimensional array of doped cadmium oxide cylinders supporting localized surface-plasmon resonances. The model predicts enhancement of the sum of the individual time-reflected modal powers by three orders of magnitude compared to a homogeneous layer of the same material operated at its epsilon-near-zero (ENZ) frequency. Our results establish a route to enhancing optical temporal reflections without relying solely on large bulk refractive-index changes.

\section{Theory of Resonance-Enhanced Time Reflection}

Exploiting material resonances to enhance time reflection at a single temporal interface seems like a natural strategy. Indeed, intrinsic material and structural resonances have been shown to substantially widen the momentum bandgaps of photonic time crystals, even under weak temporal modulations~\cite{wang2025expanding,garg2026photonic}. 
Because a stepwise photonic time crystal consists of a periodic sequence of temporal interfaces, it is natural to hypothesize that this gap widening results from stronger time reflection at its constituent interfaces. This hypothesis, however, is not generally valid. Only for the special case of a binarily and weakly modulated nondispersive photonic time crystal satisfying the quarter-period Bragg condition, the normalized momentum-gap width and the magnitude of the single-interface field-reflection coefficient become linearly proportional~\cite{asgari2024theory}. 
In resonant dispersive media, the relationship becomes more complex because temporal switching acts on internal polarization degrees of freedom and can excite multiple post-switch polariton branches, whereas the momentum gap is determined by the full Floquet evolution over an entire modulation period. Consequently, the resonance-induced widening of the momentum gap of a photonic time crystal does not by itself establish enhanced reflection at a single temporal interface (the opposite statement is also true). Demonstrating such enhancement requires a dedicated single-interface theory, which we develop below.

Furthermore, in sharp contrast to a reciprocal spatial interface, the amplitude of a time-reflected wave strongly depends on the sequence (or ``direction'') of the temporal interface. This asymmetry is already evident in the dispersionless model under temporal interface conditions that preserve the electric and magnetic fields. The electric-field reflection coefficient is \(r=\frac{1}{2}\left(1-\frac{Z_+}{Z_-}\right)=\frac{1}{2}\left(1-\sqrt{\frac{\varepsilon_-}{\varepsilon_+}}\right)\), where \(Z_-\) and \(Z_+\) are the wave impedances before and after the temporal interface, respectively, and the second equality applies to nonmagnetic media~\cite{galiffi2025electrodynamics,moussa2023observation}. Consequently, switching from a state characterized by \(Z_-\) to one characterized by \(Z_+\) is physically distinct from performing the reverse temporal transition. For positive permittivities and a fixed incident electric-field amplitude, this expression formally predicts that \(\lvert r\rvert\) grows without bound as \(\varepsilon_-\rightarrow\infty\) with \(\varepsilon_+\) fixed, whereas \(r\rightarrow1/2\) as \(\varepsilon_+\rightarrow\infty\) with \(\varepsilon_-\) fixed. It therefore suggests that time reflection can be enhanced when the initial, rather than the final, material state lies close to a resonance. It is worth noting that employing the well-known Morgenthaler relation, which describes the reflection coefficient in a nondispersive scenario under the continuity of magnetic and electric flux densities (rather than electric and magnetic fields), leads to the same intuitive conclusion [see Eq.~(15b) in Ref.~\cite{morgenthaler2003velocity}]. Namely, the initial material state should be close to a resonance to achieve significant time reflection. Because a genuine material resonance is inherently dispersive, however, this argument is only heuristic. Nevertheless, we will recover a similar conclusion below from a rigorous theory of resonant temporal interfaces. As we will show, this ``directional'' asymmetry becomes considerably richer in dispersive media because, at a fixed conserved wave vector, the material states before and after switching support multiple eigenfrequencies associated with the lower- and upper-polariton branches. Accordingly, the post-switch field cannot, in general, be represented by a single pair of time-reflected and time-refracted waves. A complete description must retain the polarization dynamics, derive the temporal interface conditions for the full field--matter state, and project the initial excitation onto all post-switch modes.

\subsection{Resonant Energy Storage in a Time-Invariant Lorentz Medium}

In this section, we analyze the cycle-averaged total energy density in a \textit{time-invariant} Lorentz medium. 
We consider a basic nonmagnetic single-oscillator Lorentz model, with $\mu_{\mathrm r}=1$, as the minimal framework for examining the role of a material resonance. The polarization density obeys~\cite[p.~43]{wooten1972optical}

\begin{equation}
\ddot{P}(t)
+
\gamma\dot{P}(t)
+
\omega_\mathrm{r}^2P(t)
=
\varepsilon_0\omega_{\mathrm p}^2E(t)\,,
\label{eq:Lorentz_time_domain}
\end{equation}

with the corresponding frequency-domain relative permittivity

\begin{equation}
\varepsilon_{\rm r}(\omega)
=\varepsilon_\infty +
\frac{\omega_{\mathrm p}^2}
{\omega_\mathrm{r}^2-\omega^2-i\gamma\omega}\,.
\label{eq:Lorentz_susceptibility}
\end{equation}
Here, we consider a fixed linear polarization in an isotropic medium and write the corresponding field and polarization components as scalar quantities.
A positive damping coefficient, $\gamma>0$, regularizes the Lorentz pole and describes a causal, passive material response. Throughout the manuscript, we use the $e^{-i \omega t}$ convention. To clarify the underlying mechanism analytically, we first consider the ideal lossless limit. The limits are taken in a specific order: we first let $\gamma\to0^+$ while keeping a finite detuning from the resonance, $\omega-\omega_\mathrm{r}\neq0$, and only then allow $\omega$ to approach $\omega_\mathrm{r}$ from within a propagation band. Thus, we consider frequencies arbitrarily close to the resonance but never evaluate the response exactly at the pole $\omega=\omega_\mathrm{r}$. We include finite damping again later to assess the robustness of the resonance-assisted enhancement against material loss.

For a monochromatic field in a low-loss dispersive medium, the standard Brillouin expression gives the cycle-averaged total energy density, including both the electromagnetic and material-oscillator contributions~\cite[p.~275]{landau2013electrodynamics}:
\begin{equation}
\overline{w}
=
\frac{1}{4} \varepsilon_0
\frac{\partial\!\left[\omega\varepsilon_{\rm r}'(\omega)\right]}
{\partial\omega}
|\tilde{E}|^2
+
\frac{1}{4}\mu_0|\tilde{H}|^2\,,
\label{eq:Brillouin_total_energy}
\end{equation}
where $\tilde{E}$ and $\tilde{H}$ denote the complex phasor field amplitudes at angular frequency $\omega$, defined as $E(t)=\Re[\tilde{E} e^{-i \omega t}]$, and $\varepsilon_{\rm r}'(\omega)$ denotes the real part of $\varepsilon_{\rm r}(\omega)$. For the lossless Lorentz model, this expression can be written equivalently as~\cite{loudon1970propagation,oughstun1988velocity}
\begin{equation}
\overline{w}
=
\frac{1}{4}\varepsilon_0\varepsilon_\infty|\tilde{E}|^2
+
\frac{1}{4}\mu_0|\tilde{H}|^2
+
\frac{\omega^2+\omega_\mathrm{r}^2}
{4\varepsilon_0\omega_{\mathrm p}^2}
|\tilde{P}|^2 =\overline{w}_{\mathrm{EM}}+ \overline{w}_{\mathrm{P}}\,,
\label{eq:Lorentz_average_total_energy}
\end{equation}
where $\tilde{P}$ is the complex phasor amplitude of the polarization density. The last term represents the cycle-averaged kinetic and potential energies associated with the material polarization, i.e., $\overline{w}_{\mathrm{P}}=\overline{w}_{\mathrm{kin}}+\overline{w}_{\mathrm{pot}}$. Using
\begin{equation}
\tilde{P}
=
\varepsilon_0
\frac{\omega_{\mathrm p}^2}
{\omega_\mathrm{r}^2-\omega^2}
\tilde{E}\,,
\label{eq:lossless_polarization_amplitude}
\end{equation}
the polarization contribution becomes
\begin{equation}
\overline{w}_{\mathrm P}
=
\frac{\omega^2+\omega_\mathrm{r}^2}
{4\varepsilon_0\omega_{\mathrm p}^2}
|\tilde{P}|^2
=
\frac{\varepsilon_0\omega_{\mathrm p}^2}{4}
\frac{\omega^2+\omega_\mathrm{r}^2}
{\left(\omega_\mathrm{r}^2-\omega^2\right)^2}
|\tilde{E}|^2\,.
\label{eq:average_polarization_energy}
\end{equation}

For a fixed nonzero electric-field amplitude, Eq.~(\ref{eq:average_polarization_energy}) increases strongly as the excitation frequency approaches the material resonance and becomes unbounded in the idealized lossless resonant limit. The resonant state, therefore, contains a large amount of energy in the material polarization, in addition to the electromagnetic-field energy. In any physical system, this polarization energy is bounded by finite damping, finite excitation bandwidth, nonlinear saturation, and other nonideal effects. The lossless limit considered here serves as a reference for isolating the underlying enhancement mechanism.

The role of the resonance in enhancing temporal reflection can now be stated more precisely. When the incident wave is tuned close to the material resonance, the polarization response, and hence the material-oscillator contribution to the pre-switch energy density, can become large. The abrupt change in the material parameters then performs work on this strongly polarized field-matter state, either injecting or extracting energy, depending on the switching protocol. The temporal interface conditions determine how the resulting state is projected onto the eigenmodes of the post-switch medium. Only the portion projected onto the backward-propagating modes contributes to temporal reflection. The remaining energy is carried by the forward-propagating modes and, when damping is present, partly transferred to dissipative channels. Thus, the enhancement is governed jointly by the resonantly enhanced polarization energy, the work exchanged with the modulation, and the modal projection at the temporal interface. The following sections derive these contributions explicitly.

\subsection{Energy Exchange and Modal Redistribution at a Temporal Interface}\label{energyexch}

\begin{figure}[t]
    \centering
    \includegraphics[width=0.6\linewidth]{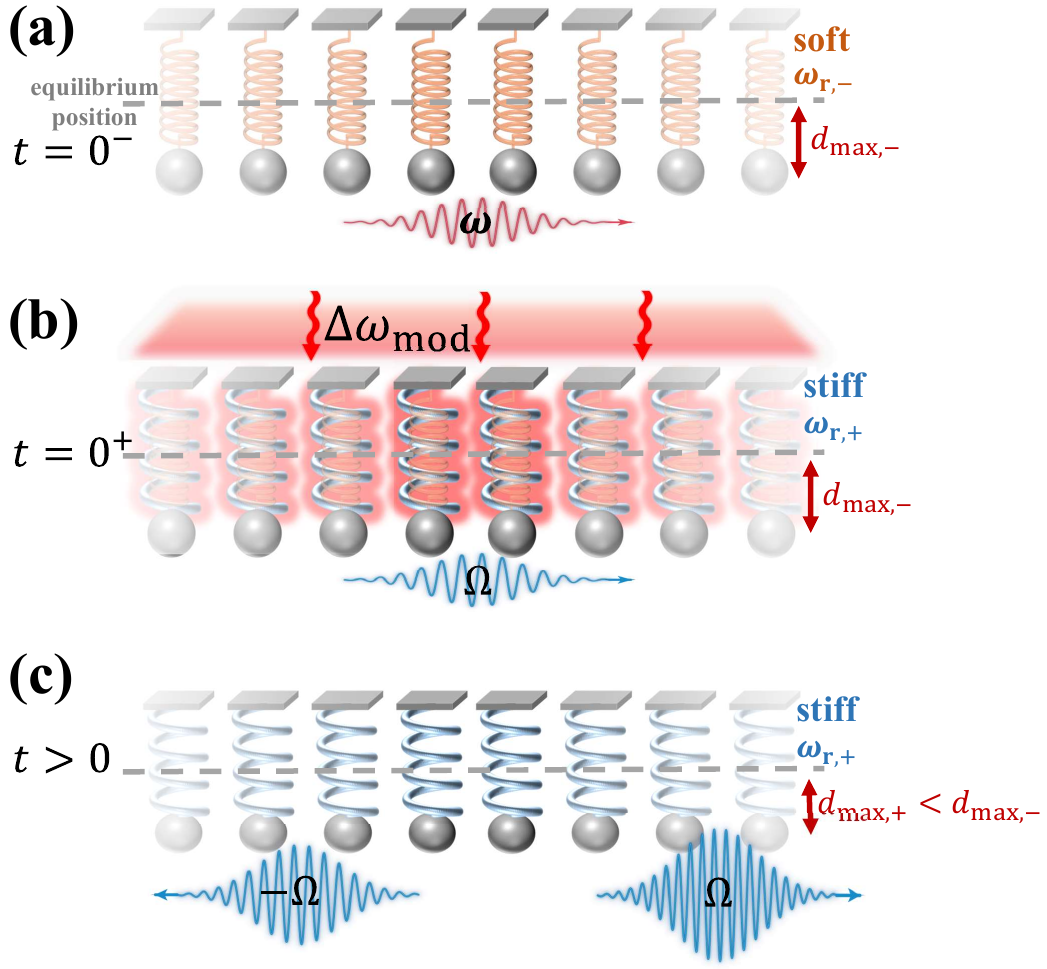}
    \caption{Conceptual mechanical analogy of resonance-assisted time-reflection enhancement. The repeated oscillators represent copies of the local field--matter state at a single spatial phase and are therefore shown oscillating in phase. (a) Before switching (\(t=0^-\)), a wave of frequency \(\omega\) excites strong polarization oscillations near the initial resonance \(\omega_{\mathrm{r},-}\), storing substantial energy in the coupled field--matter state. The soft springs represent the lower initial resonance frequency, and their displacement from the equilibrium position represents the polarization \(P\). (b) At \(t=0^+\), a spatially uniform modulation abruptly increases the resonance frequency to \(\omega_{\mathrm{r},+}>\omega_{\mathrm{r},-}\), represented by stiffer springs. Because \(P\) and \(\dot P\) remain continuous, the increased stiffness instantaneously raises the polarization potential energy by \(\Delta w_{\mathrm{mod}}>0\), supplied by the modulation mechanism. (c) This energy, together with the initially stored energy, is subsequently redistributed among the post-switch polariton modes, generating enhanced time-reflected (\(-\Omega\)) and time-refracted (\(+\Omega\)) waves.}
    \label{fig:concept}
\end{figure}

Having characterized the frequency-dependent energy distribution in a time-invariant Lorentz medium, we now examine how the total energy changes when the material resonant frequency $\omega_\mathrm{r}$ and the damping coefficient $\gamma$ are switched abruptly in time (much faster than the wave cycle). 
Using Poynting's theorem for the Lorentz medium described by Eq.~(\ref{eq:Lorentz_time_domain}), in which the resonance frequency $\omega_{\mathrm r}(t)$ and the damping coefficient $\gamma(t)$ vary in time while $\varepsilon_\infty$ and $\omega_{\mathrm p}$ remain constant, we obtain the following local time-domain energy-balance equation (see Supplementary Section~1):
\begin{equation}
\frac{\partial w_{\mathrm{tot}}(\mathbf r,t)}{\partial t}
+
\nabla\cdot\mathbf S(\mathbf r,t)
=
\frac{
\mathbf P(\mathbf r,t)\cdot\mathbf P(\mathbf r,t)
}{
2\varepsilon_0\omega_{\mathrm p}^2
}
\frac{\partial\omega_{\mathrm r}^2(t)}{\partial t}
-
\frac{
\gamma(t)\,
\dot{\mathbf P}(\mathbf r,t)
\cdot
\dot{\mathbf P}(\mathbf r,t)
}{
\varepsilon_0\omega_{\mathrm p}^2
}\,,
\label{eq:time_domain_energy_balance}
\end{equation}
where
\begin{equation}
w_{\mathrm{tot}}(\mathbf r,t)
=
\frac{\varepsilon_0\varepsilon_\infty}{2}
\mathbf E(\mathbf r,t)\cdot\mathbf E(\mathbf r,t)
+
\frac{\mu_0}{2}
\mathbf H(\mathbf r,t)\cdot\mathbf H(\mathbf r,t)
+
\frac{
\dot{\mathbf P}(\mathbf r,t)
\cdot
\dot{\mathbf P}(\mathbf r,t)
+
\omega_{\mathrm r}^2(t)
\mathbf P(\mathbf r,t)\cdot\mathbf P(\mathbf r,t)
}{
2\varepsilon_0\omega_{\mathrm p}^2
}
\label{eq:supp_total_energy_density}
\end{equation}
is the instantaneous total energy density, including the electromagnetic and Lorentz-oscillator contributions. It is related to the cycle-averaged expression in Eq.~(\ref{eq:Lorentz_average_total_energy}). Here, $\dot{\mathbf P}(\mathbf r,t)\equiv\frac{\partial\mathbf P(\mathbf r,t)}{\partial t}$, and the instantaneous Poynting vector is
\begin{equation}
\mathbf S(\mathbf r,t)
=
\mathbf E(\mathbf r,t)\times\mathbf H(\mathbf r,t)\,.
\end{equation}
Because we consider an isotropic bulk medium and a fixed polarization direction, the vector polarization is projected onto that direction. Thus, for brevity, in the remainder of the manuscript and throughout the Supplementary Material, we use the scalar polarization variable $P(t)$ and omit the spatial argument $\mathbf r$ from the notation.

The first term on the right-hand side of (\ref{eq:time_domain_energy_balance}) represents the power supplied by the temporal modulation, whereas the second term describes material dissipation. In the lossless limit considered here, \(\gamma\rightarrow0^+\), the first (modulation) term is the only local source of energy.

We consider an abrupt and spatially uniform change in the resonance frequency at \(t=0\),
\begin{equation}
\omega_\mathrm{r}^2(t)
=
\omega_{\mathrm{r},-}^2
+
\left(
\omega_{\mathrm{r},+}^2-\omega_{\mathrm{r},-}^2
\right)
U(t)\,,
\label{eq:abrupt_resonance_switch}
\end{equation}
where \(U(t)\) is the unit step function and $\omega_{\mathrm{r},-}$ and $\omega_{\mathrm{r},+}$ represent the frequency before and after the jump, respectively. Integrating Eq.~(\ref{eq:time_domain_energy_balance}) across an infinitesimal interval containing the temporal interface gives (see Supplementary Section~1)
\begin{equation}
w_{\mathrm{tot}}(0^+)-w_{\mathrm{tot}}(0^-)
=
\frac{\omega_{\mathrm{r},+}^2-\omega_{\mathrm{r},-}^2}
{2\varepsilon_0\omega_{\mathrm p}^2}
P^2(0)=\Delta w_{\mathrm{mod}}\,.
\label{eq:interfacial_modulation_work}
\end{equation}
Equation~(\ref{eq:interfacial_modulation_work}) gives the exact work density performed by the temporal modulation. The change in the total energy on the left-hand side of (\ref{eq:interfacial_modulation_work}) follows from the temporal interface conditions. When \(\varepsilon_\infty\) and \(\omega_{\mathrm p}\) remain unchanged and no impulsive sources are applied~\cite{solis2021time,bakunov2021constitutive,bakunov2021light},
\begin{equation}
D(0^+)=D(0^-)\,,
\qquad
B(0^+)=B(0^-)\,,
\qquad
P(0^+)=P(0^-)\,,
\qquad
\dot{P}(0^+)=\dot{P}(0^-)\,.
\label{eq:complete_temporal_boundary_conditions}
\end{equation}
Since \(D=\varepsilon_0\varepsilon_\infty E+P\), the continuity of \(D\) and \(P\) also implies that \(E\) is continuous. Likewise, \(H\) is continuous when the permeability is unchanged. Here, these continuity conditions apply to the total instantaneous electromagnetic fields and the total polarizations, rather than to those of a single polaritonic mode. The electromagnetic energy and polarization kinetic energy, therefore, do not change instantaneously, as is seen from (\ref{eq:supp_total_energy_density}). The entire discontinuity in the total energy originates from the change in the polarization \textit{potential} energy caused by the abrupt variation of the resonance frequency.

Equation~(\ref{eq:interfacial_modulation_work}) makes both the sign and magnitude of the energy exchange transparent. An upward shift, $\omega_{\mathrm{r},+}>\omega_{\mathrm{r},-}$, increases the restoring stiffness of the material oscillator and, because $P$ remains continuous during the switch, instantaneously raises its potential energy. This additional energy is supplied by the modulation mechanism so that $\Delta w_{\mathrm{mod}}>0$. Conversely, a downward shift reduces the restoring stiffness and extracts energy from the system. For a fixed resonance shift, the exchanged energy is proportional to $P^2(0)$. Locally, it is therefore largest when the switch occurs at maximum polarization magnitude and vanishes when $P(0)=0$. A direct mechanical analog, illustrated in Fig.~\ref{fig:concept}, is a heavy mass attached to a spring. Suppose that, at the instant of maximum extension, when the mass is momentarily at rest, the spring is abruptly changed from a soft, rubber-like spring (Fig.~\ref{fig:concept}(a)) to a stiff, metal-like one without changing its instantaneous extension (the mechanical analog of $P$), as shown in Fig.~\ref{fig:concept}(b). Its potential energy $U=Kx^2/2$ then increases by $\Delta U=(K_+-K_-)x^2/2$, and the mechanism that increases the spring constant $K$ must supply this energy. Because $\Delta U\propto x^2$, switching the stiffness at maximum extension produces the largest energy increase, precisely mirroring the enhancement obtained by changing $\omega_{\mathrm r}^2$ when $P^2$ is maximal.

Equation~(\ref{eq:interfacial_modulation_work}) shows that the local work density depends on the instantaneous polarization and is therefore sensitive to the phase at which the temporal switching occurs. For the monochromatic incident plane wave considered here, $P(\mathbf{r},0)=|\tilde{P}|\cos\phi(\mathbf{r})$, where $\phi$ spans $2\pi$ over one spatial wavelength. Consequently, the local work density also varies with position. To characterize the macroscopic energy exchange, we average the work density over one wavelength, which is equivalent to averaging over $\phi$, and denote the result by $\left\langle\Delta w_{\mathrm{mod}}\right\rangle_{\phi}$. This average does not vanish because the modulation work depends quadratically on the polarization, so opposite signs of $P$ make equal contributions. Specifically, although $\left\langle P(0)\right\rangle_{\phi}=0$, one has $\left\langle P^2(0)\right\rangle_{\phi}=|\tilde{P}|^2/2$. Therefore,
\begin{equation}
\left\langle\Delta w_{\mathrm{mod}}\right\rangle_{\phi}
=
\frac{\omega_{\mathrm{r},+}^2-\omega_{\mathrm{r},-}^2}
{4\varepsilon_0\omega_{\mathrm p}^2}
|\tilde{P}|^2\,.
\label{eq:phase_averaged_modulation_work}
\end{equation}

To quantify how the initial resonance affects the energy distribution at the temporal interface, we compare monochromatic incident waves at different frequencies while keeping both material states, and hence the resonance shift $\omega_{\mathrm{r},+}-\omega_{\mathrm{r},-}$, fixed. We also keep the incident electric-field amplitude $|\tilde{E}|$ fixed, so that the frequency dependence reflects only the resonant response of the medium. We begin with the energy stored immediately before switching. In the lossless initial medium, whose relative permittivity, as given by (\ref{eq:Lorentz_susceptibility}), equals
$\varepsilon_{\mathrm{r},-}(\omega)=\varepsilon_\infty+\omega_{\mathrm p}^2/(\omega_{\mathrm{r},-}^2-\omega^2)$, the cycle-averaged electromagnetic-field energy is
\begin{equation}
\begin{aligned}
\overline{w}_{\mathrm{EM}}
&=
\frac{1}{4}\varepsilon_0\varepsilon_\infty|\tilde{E}|^2
+
\frac{1}{4}\mu_0|\tilde{H}|^2
\\
&=
\frac{1}{4}\varepsilon_0
\left[
\varepsilon_\infty+\varepsilon_{\mathrm{r},-}(\omega)
\right]
|\tilde{E}|^2
\\
&=
\frac{1}{4}\varepsilon_0
\left[
2\varepsilon_\infty
+
\frac{\omega_{\mathrm p}^2}
{\omega_{\mathrm{r},-}^2-\omega^2}
\right]
|\tilde{E}|^2\,.
\end{aligned}
\label{eq:pre_switch_em_energy}
\end{equation}
Thus, for a fixed $|\tilde{E}|$, the resonant increase of $\overline{w}_{\mathrm{EM}}$ originates from the magnetic-field contribution, which grows as the wave number sharply increases near the resonance (or, equivalently, as the wave impedance significantly decreases).

The material-oscillator energy exhibits an even stronger resonant dependence. Using Eqs.~(\ref{eq:lossless_polarization_amplitude})--(\ref{eq:average_polarization_energy}) and separating the material energy into kinetic and potential contributions gives
\begin{equation}
\begin{aligned}
\overline{w}_{\mathrm{kin}}
&=
\frac{\omega^2}{4\varepsilon_0\omega_{\mathrm p}^2}
|\tilde{P}|^2
=
\frac{\varepsilon_0\omega_{\mathrm p}^2}{4}
\frac{\omega^2}
{\left(\omega_{\mathrm{r},-}^2-\omega^2\right)^2}
|\tilde{E}|^2\,,
\\
\overline{w}_{\mathrm{pot}}
&=
\frac{\omega_{\mathrm{r},-}^2}
{4\varepsilon_0\omega_{\mathrm p}^2}
|\tilde{P}|^2
=
\frac{\varepsilon_0\omega_{\mathrm p}^2}{4}
\frac{\omega_{\mathrm{r},-}^2}
{\left(\omega_{\mathrm{r},-}^2-\omega^2\right)^2}
|\tilde{E}|^2\,.
\end{aligned}
\label{eq:pre_switch_polarization_components}
\end{equation}
These expressions reveal a clear energy hierarchy as $\omega$ approaches $\omega_{\mathrm{r},-}$ from the lower propagation band. The resonant part of the electromagnetic-field energy scales as
$\left(\omega_{\mathrm{r},-}^2-\omega^2\right)^{-1}$, whereas both components of the polarization energy scale as
$\left(\omega_{\mathrm{r},-}^2-\omega^2\right)^{-2}$. Consequently, the polarization energy increasingly dominates the total pre-switch energy. Because the phase-averaged modulation work in Eq.~(\ref{eq:phase_averaged_modulation_work}) is also proportional to $|\tilde{P}|^2$, its magnitude has the same inverse-square resonant enhancement. Thus, for the same fixed resonance shift, bringing the incident wave closer to the initial resonance simultaneously increases both the energy already stored in the material oscillator and the magnitude of the energy taken from the modulation mechanism. The sign of the latter remains determined by the direction of the resonance frequency shift: resonance proximity enhances both energy injection for an upward shift and energy extraction for a downward shift.

Before the temporal interface, we assume that the field--matter state is a single monochromatic eigenmode of the initial Lorentz medium. Its cycle-averaged total energy density, denoted by $\overline{w}_{-}$, is obtained from Eq.~(\ref{eq:Lorentz_average_total_energy}) by setting $\omega_{\mathrm r}=\omega_{\mathrm{r},-}$ and using the pre-switch phasor amplitudes $(\tilde{E}_{-},\tilde{H}_{-},\tilde{P}_{-})$.
At the switching instant, the temporal interface conditions in Eq.~(\ref{eq:complete_temporal_boundary_conditions}) preserve the instantaneous field--matter state, but the eigenmodes supported by the medium change. Consequently, the inherited state is generally not a single eigenmode of the post-switch medium and must instead be expanded over its complete set of eigenmodes. Because the temporal modulation is spatially uniform, the wave vector $k$ is conserved. At this fixed wave vector, the post-switch Lorentz medium supports lower- and upper-polariton branches, each with positive- and negative-frequency components corresponding, respectively, to forward- and backward-propagating waves. We use $\nu$ to label all these post-switch modes, whose amplitudes are determined by the temporal interface conditions~\cite{solis2021time}.
Averaging over a time interval long compared with the beat periods between the generated modes eliminates the oscillatory interference terms. The resulting post-switch total energy density is therefore the sum of the individual modal contributions:
\begin{equation}
\overline{w}_{+}
=
\sum_{\nu}
\left[
\frac{1}{4}\varepsilon_0\varepsilon_\infty
|\tilde{E}_\nu|^2
+
\frac{1}{4}\mu_0
|\tilde{H}_\nu|^2
+
\frac{\Omega_\nu^2}
{4\varepsilon_0\omega_{\mathrm p}^2}
|\tilde{P}_\nu|^2
+
\frac{\omega_{\mathrm{r},+}^2}
{4\varepsilon_0\omega_{\mathrm p}^2}
|\tilde{P}_\nu|^2
\right]\,.
\label{eq:post_switch_modal_energy}
\end{equation}
Here, $\Omega_\nu$ is the eigenfrequency of the $\nu$th post-switch mode, and $(\tilde{E}_\nu,\tilde{H}_\nu,\tilde{P}_\nu)$ are its complex phasor amplitudes. The four terms in Eq.~(\ref{eq:post_switch_modal_energy}) are, respectively, the electric-field, magnetic-field, polarization-kinetic, and polarization-potential energy densities. The sum extends over all generated modes on both polariton branches and in both propagation directions.

\begin{figure}[tb]
    \centering
    \includegraphics[width=0.98\linewidth]{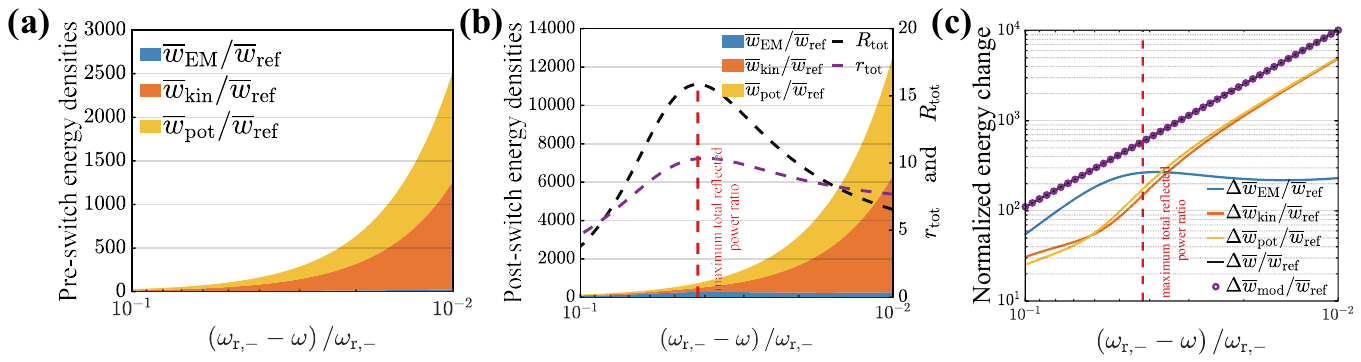}
    \caption{Cycle-averaged energy density redistribution across the temporal interface at which the resonance frequency abruptly increases from $\omega_{\mathrm{r},-}$ to $\omega_{\mathrm{r},+}=M_{\rm r}\omega_{\mathrm{r},-}$ with $M_{\rm r}=3$ for a fixed incident electric-field amplitude. All energy densities are normalized to the frequency-independent reference energy $\overline{w}_{\mathrm{ref}}$ and plotted as functions of the normalized frequency detuning. (a) Pre-switch energy distribution, including the electromagnetic, polarization-kinetic, and polarization-potential contributions. (b) Post-switch energy distribution summed over all generated lower- and upper-polariton modes. The black dashed curve shows the total reflected power ratio, while the purple dashed curve shows the sum of the magnitudes of the upper- and lower-mode electric-field reflection coefficients, both read from the right axis. (c) Normalized changes in the electromagnetic, polarization-kinetic, and polarization-potential energy densities, together with the total energy density change. The purple open circles represent the phase-averaged work performed by the temporal modulation. In (b) and (c), the red dashed vertical line marks the incident frequency at which the reflected power ratio reaches its maximum. This occurs at $\omega/\omega_{\mathrm{r},-}=0.958$, at which  $\Delta\overline{w}_{\mathrm{EM}}/\overline{w}_{\mathrm{ref}}=2.649\times10^{2}$, $\Delta\overline{w}_{\mathrm{kin}}/\overline{w}_{\mathrm{ref}}=1.49\times10^{2}$, $\Delta\overline{w}_{\mathrm{pot}}/\overline{w}_{\mathrm{ref}}=1.746\times10^{2}$, and $\Delta\overline{w}/\overline{w}_{\mathrm{ref}}=5.8859\times10^{2}$.}
    \label{fig:energy1}
\end{figure}

To illustrate this energy redistribution quantitatively, we consider an abrupt increase of the resonance frequency from 
$\omega_{\mathrm{r},-}$ to $\omega_{\mathrm{r},+}=M_{\rm r}\omega_{\mathrm{r},-}$, where $M_{\rm r}$ is the modulation factor for the resonance frequency. The relation between the plasma frequency and the resonance frequency before the temporal jump is given by the effective oscillator strength parameter $N = \omega_{{\rm p}}/\omega_{{\rm r,-}}$. Here, we choose $M_{\rm r}=3$ and $N=1$.
We vary the incident frequency $\omega$ while keeping the magnitude of the incident electric-field amplitude, $|\tilde{E}_{-}|$, fixed. To place all energy contributions on a common, frequency-independent scale, we normalize them by
$\overline{w}_{\mathrm{ref}}=\frac{1}{2}\varepsilon_0\varepsilon_\infty|\tilde{E}_{-}|^2$. This reference corresponds to the cycle-averaged electromagnetic energy density of a plane wave with the same electric-field amplitude propagating in the nondispersive background medium $\varepsilon_\infty$.
Figure~\ref{fig:energy1}(a) decomposes the total pre-switch energy density $\overline{w}_{-}$ into its electromagnetic, polarization-kinetic, and polarization-potential contributions. As $\omega$ approaches the initial resonance $\omega_{\mathrm{r},-}$ from the lower propagation band, the polarization amplitude increases as
$|\tilde{P}_{-}|\propto|\omega_{\mathrm{r},-}^2-\omega^2|^{-1}$. Consequently, both polarization-energy contributions grow as
$|\omega_{\mathrm{r},-}^2-\omega^2|^{-2}$ and rapidly dominate the pre-switch energy, consistent with Eq.~(\ref{eq:pre_switch_polarization_components}). Moreover, because
$\overline{w}_{\mathrm{kin}}/\overline{w}_{\mathrm{pot}}=\omega^2/\omega_{\mathrm{r},-}^2$, the kinetic and potential contributions become nearly equal close to the resonance.

Figure~\ref{fig:energy1}(b) shows how the total post-switch energy is distributed among the different degrees of freedom of all generated modes. The electromagnetic-field energy density associated with the negative-frequency modes is given by $ 
\overline{w}_{\mathrm{EM,r}}
=
\overline{w}_{\mathrm{EM,r,L}}
+
\overline{w}_{\mathrm{EM,r,U}},
$
where subscripts 'L' and 'U' denote the lower- and upper-polariton branches, respectively. The same panel also shows on the right axis the sum of the upper- and lower-mode normalized reflected electric fields $r_{\rm tot}$ and the total reflected power ratio $R_{\rm tot}=|\overline{S}_{r}|/\overline{S}_{-}$ defined as the ratio of the corresponding Poynting vector magnitudes. 
Importantly, both $r_{\rm tot}$ and $R_{\mathrm{tot}}$ exhibit pronounced maxima at nearly the same finite red detuning from the initial resonance frequency. The reflected-power maximum corresponds to a conserved wave number near the avoided crossing of the post-switch polariton branches, rather than to the limit $\omega\to\omega_{\mathrm{r},-}$. This connection is discussed in more detail in Sec.~\ref{comparingDMT}. Thus, maximizing the initially stored polarization energy \textit{alone} does not necessarily maximize the reflected electromagnetic field or the associated energy flux. The optimum also depends on the modulation-induced energy transfer, the projection of the inherited field--matter state onto the negative-frequency post-switch modes, and the electromagnetic content and impedance of those modes.
\begin{figure}[t]
    \centering
    \includegraphics[width=0.99\linewidth]{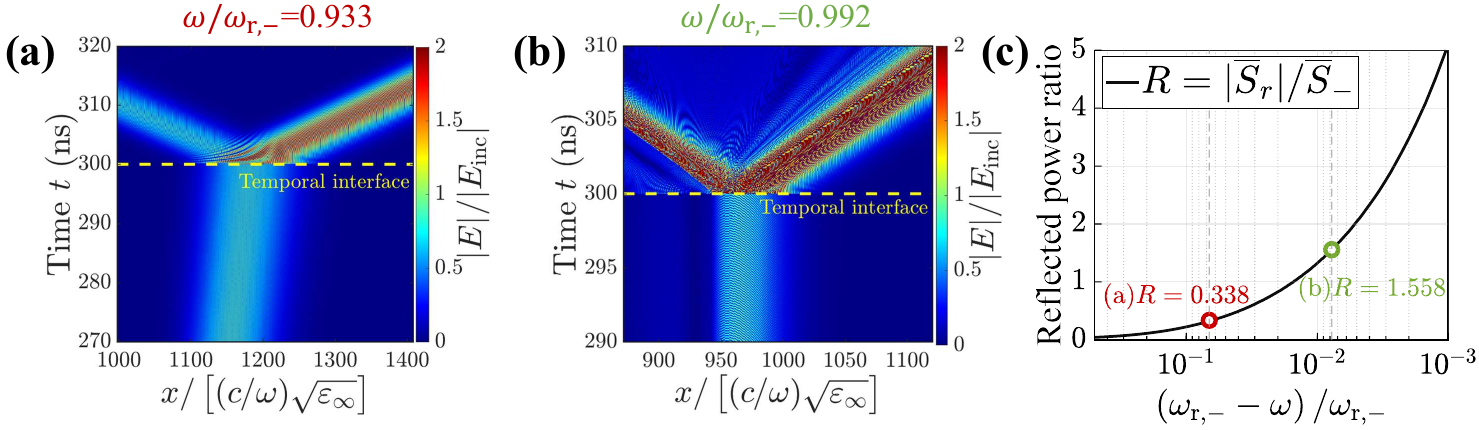}
    \caption{FDTD validation of resonance-enhanced time reflection for a Lorentz-to-vacuum temporal interface. (a)--(b) Spatiotemporal evolution of the normalized electric-field magnitude for \(\omega/\omega_{\mathrm r,-}=0.933\) and \(0.992\), respectively. For each case, \(E_{\mathrm{inc}}\) denotes the peak amplitude of the incident electric-field at \(t=300\,\mathrm{ns}\), immediately before switching. The green dashed lines mark the temporal interface at \(t=300\,\mathrm{ns}\), where the incident pulse splits into time-reflected and time-refracted pulses. (c) Analytical time-reflected power ratio, \(R=\lvert\overline{S}_{r}\rvert/\overline{S}_{-}\), as a function of the normalized detuning \((\omega_{\mathrm r,-}-\omega)/\omega_{\mathrm r,-}\). The red and green circles show the corresponding FDTD results extracted from (a) and (b), yielding \(R=0.338\) and \(R=1.558\), respectively.
}
    \label{fig:FDTD1}
\end{figure}

The change of different energy density components across the temporal interface is quantified in Fig.~\ref{fig:energy1}(c). Here, $\Delta\overline{w}=\overline{w}_{+}-\overline{w}_{-}$ is the total net energy increase of the wave-matter state.   Importantly, one can observe the complete overlap between  $\Delta\overline{w}$ and the phase-averaged work $\left\langle\Delta w_{\mathrm{mod}}\right\rangle_{\phi}$ produced by the modulation mechanism during the temporal interface. This agreement connects the \textit{instantaneous} time-domain energy balance at the temporal interface with the independent modal calculation of the \textit{time-averaged} post-switch energy.
The equality $\Delta\overline{w} = \left\langle\Delta w_{\mathrm{mod}}\right\rangle_{\phi}$ does not imply that the modulation work is transferred directly to the electromagnetic field. At the switching instant, the continuity of $E$, $H$, $P$, and $\dot{P}$ leaves the electromagnetic and polarization-kinetic energies unchanged. For the upward resonance shift considered here, the entire instantaneous energy increase initially appears in the polarization-potential energy. The subsequent coupled field--matter dynamics redistribute this energy among the electromagnetic, polarization-kinetic, and polarization-potential sectors. This redistribution can be seen in Fig.~\ref{fig:energy1}(c). 
The positive dominant value of $\Delta\overline{w}_{\mathrm{EM}}$ demonstrates that a substantial fraction of the energy initially deposited in the potential-polarization channel is subsequently converted into electromagnetic energy, thereby feeding the time-reflected waves, as schematically illustrated in Fig.~\ref{fig:concept}(c).

\subsection{Numerical Validation}

We next validate the resonance-assisted time-reflection enhancement mechanism by comparing analytical scattering coefficients with one-dimensional FDTD simulations. We consider a bulk medium described by the single-resonance Lorentz permittivity given by (\ref{eq:Lorentz_susceptibility}) with \(\gamma=0\). Details of the modal calculations, FDTD implementation, and extraction of the scattering amplitudes are provided in the Methods section.

\begin{figure}[t]
    \centering
    \includegraphics[width=0.9\linewidth]{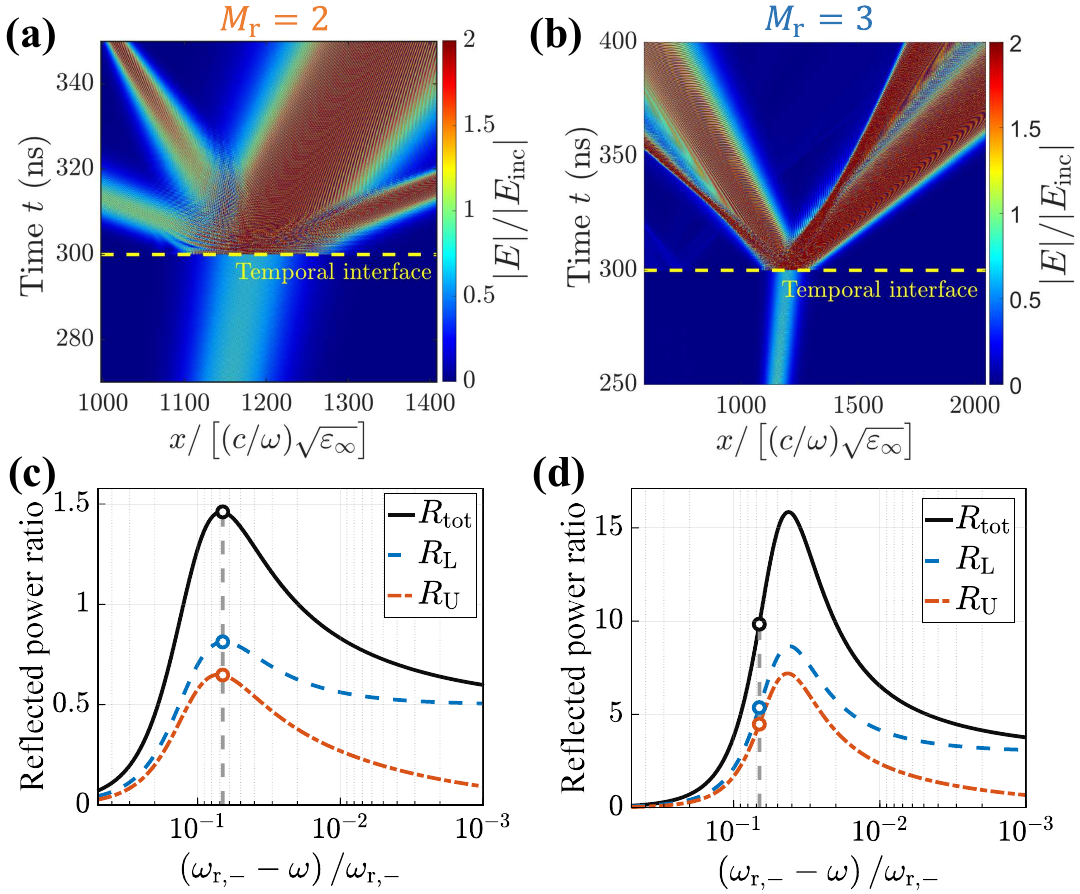}
    \caption{FDTD validation of resonance-enhanced time reflection for a Lorentz-to-Lorentz temporal interface. (a) and (b) Spatiotemporal evolution of the normalized electric-field magnitude $\lvert E\rvert/\lvert E_{\mathrm{inc}}\rvert$ for $M_{\rm r}=2$ and $M_{\rm r}=3$, respectively. The green dashed lines indicate the temporal interface. After switching, the incident pulse separates into two time-reflected and two time-refracted pulses associated with the two post-switch polariton branches. (c) and (d) Analytically calculated total, upper- and lower-branch reflected power ratios as functions of the normalized detuning $(\omega_{\mathrm r,-}-\omega)/\omega_{\mathrm r,-}$ for $M_{\rm r}=2$ and $M_{\rm r}=3$, respectively. The black solid curves show the total reflected power ratio $R_{\mathrm{tot}}$, while the blue and orange dashed curves show the contributions from the lower and upper polariton branches, $R_{L}$ and $R_{U}$, respectively. Open circles denote the corresponding FDTD results at the selected incident frequency.}
    \label{fig:FDTD2}
\end{figure}

We first examine the limiting case in which the initial Lorentz response is completely removed at the temporal interface, so that \(\omega_{\mathrm p,+}=0\) and the post-switch medium is vacuum. Because the vacuum state supports only one positive- and one negative-frequency mode at the conserved wave number, the temporal scattering consists of a single time-refracted wave and a single time-reflected wave. Under the electric- and magnetic-field-preserving temporal-interface convention adopted for this limiting switch, their electric-field and power time-reflection coefficients are~(see Supplementary Section~2)
\begin{equation}
\begin{aligned}
r
&=
\frac{1}{2}
\left[
1-\sqrt{\varepsilon_{\mathrm r,-}(\omega)}
\right]
=
\frac{1}{2}
\left[
1-
\sqrt{
1+
\frac{\omega_{\mathrm p,-}^{2}}
{\omega_{\mathrm r,-}^{2}-\omega^{2}}
}
\right]\,,
\\
R
&=
|r|^{2}\frac{\omega}{\Omega}\,,
\\
\Omega
&=
c k
=
\omega\sqrt{\varepsilon_{\mathrm r,-}(\omega)}
=
\omega
\sqrt{
1+
\frac{\omega_{\mathrm p,-}^{2}}
{\omega_{\mathrm r,-}^{2}-\omega^{2}}
}\,.
\end{aligned}
\label{eq:vacuum}
\end{equation}
Here, \(\omega\) and $\Omega$ are the incident frequency and frequency after switching, respectively. The scattering coefficients are normalized to the incident electric-field amplitude. The incident frequency is always smaller than the resonance frequency, so we always excite the lower polaritonic branch mode. Figures~\ref{fig:FDTD1}(a) and \ref{fig:FDTD1}(b) show the simulated field evolution in the space-time coordinate system for \(\omega/\omega_{\mathrm r,-}=0.933\) and \(0.992\), respectively. Following the temporal switch, the incident pulse separates into counterpropagating time-refracted and time-reflected pulses. 
The analytical time-reflected power coefficient from (\ref{eq:vacuum}) is compared with the FDTD results in Fig.~\ref{fig:FDTD1}(c), revealing excellent agreement. 
As is seen, 
the reflected power ratio $R=\lvert\overline{S}_{\rm r}\rvert/\overline{S}_{-}$ increases from 0.338 at $\omega/\omega_{\mathrm r,-}=0.933$ to 1.558 at $\omega/\omega_{\mathrm r,-}=0.992$. The latter value exceeds unity, indicating that the reflected Poynting flux becomes larger than the incident Poynting flux near resonance. A zoomed-in view of the spatiotemporal evolution shown in Figs.~\ref{fig:FDTD1}(a) and (b) can be found in Supplementary Section~2.
By comparison, the reflected amplitude reaches only $R \approx 0.03$ in the off-resonant limit. 
Interestingly, as shown in Fig.~\ref{fig:FDTD1}(c), for the Lorentz-to-vacuum temporal interface, the time-reflected power ratio $R$ increases monotonically as the incident frequency approaches the initial resonance, in contrast to the Lorentz-to-Lorentz case shown in Fig.~\ref{fig:energy1}(b), where a pronounced maximum occurs at a finite distance from resonance. This difference originates from the distinct post-switch modal structures of the two systems. In the Lorentz-to-vacuum case, the reflected field is carried by a single backward vacuum mode, and its reflected power increases relatively slowly as the initial resonance is approached. In the Lorentz-to-Lorentz case, however, both the lower- and upper-polariton backward branches contribute to the reflected power. Their different frequency-dependent growth rates lead to a much more rapid enhancement of the total reflected power and produce a pronounced maximum at a finite frequency below resonance. Consequently, even at $\omega/\omega_{\mathrm{r},-}=0.992$, very close to resonance, the Lorentz-to-vacuum case gives only $R=1.558$, substantially smaller than the maximum reflected-power ratio obtained for the Lorentz-to-Lorentz switching.

We next consider Lorentz-to-Lorentz time interfaces in which the resonance frequency is increased while the oscillator strength $\omega_{\rm p}$ remains unchanged. Specifically, we consider scenarios with $M_{\rm r}=2$ and $M_{\rm r}=3$. We select these comparatively large shifts to make the enhancement and the resulting multimode temporal scattering particularly clear. Unlike vacuum, the post-switch Lorentz medium supports lower- and upper-polariton branches at the conserved wave number. Each branch contributes positive- and negative-frequency components, producing two time-refracted waves, characterized by power coefficients \(T_{\rm L}\) and \(T_{\rm U}\), and two time-reflected waves, characterized by \(R_{\rm L}\) and \(R_{\rm U}\).
The four generated wave packets can be directly identified in the simulated field evolutions in Figs.~\ref{fig:FDTD2}(a) and \ref{fig:FDTD2}(b). Due to the finite spectral bandwidth of the incident pulse, each post-switch polariton branch experiences group-velocity dispersion, characterized by $k''(\omega)=\partial^2 k/\partial\omega^2$. Different spectral components, therefore, propagate with different group velocities, leading to temporal and spatial reshaping of the wave packets and consequently increasing their widths along the $x$-direction during propagation. 

To further quantify the time-reflected waves, Figs.~\ref{fig:FDTD2}(c) and \ref{fig:FDTD2}(d) show the total reflected-power ratio $R_{\mathrm{tot}}=R_{\mathrm L}+R_{\mathrm U}$ (black solid curves), together with the lower- and upper-branch contributions, $R_{\mathrm L}$ and $R_{\mathrm U}$ (blue and orange dashed curves), for $M_{\mathrm r}=2$ and $3$, respectively. The reflected electric-field coefficients and branch-resolved power ratios are given by (see Ref.~\cite{solis2021time} and Supplementary Section~2)
\begin{equation}
\begin{aligned}
r_j
&=
\frac{\Omega_j}{2\omega}
\left(\frac{\Omega_j}{\omega}-1\right)
\frac{\Omega_{\bar{j}}^{2}-\omega^{2}}
     {\Omega_{\bar{j}}^{2}-\Omega_j^{2}}\,,
\\[4pt]
R_j
&=
|r_j|^{2}\frac{\omega}{\Omega_j}\,,
\qquad j=\mathrm L,\mathrm U\,.
\end{aligned}
\label{eq:modal_reflection_coefficients}
\end{equation}
Here, $\omega$ is the incident frequency, $\Omega_{\mathrm L}$ and $\Omega_{\mathrm U}$ are the post-switch lower- and upper-branch frequencies, and $\bar{j}$ denotes the branch other than $j$, i.e., ${\rm \bar{L} = U}$ and ${\rm \bar{U} = L}$. The coefficients $r_j$ and $R_j$ are normalized to the incident electric-field amplitude and Poynting flux, respectively. The analytical curves agree closely with the FDTD results. In the low-frequency, nonresonant regime, the reflected-power ratios remain below $0.1$ but increase sharply as the incident frequency approaches the initial resonance, confirming the enhancement of both reflected branches. The ratios exhibit enhancements exceeding a factor of $150$ relative to the nonresonant case and reach their maxima at finite red detunings from $\omega_{\mathrm{r},-}$ rather than at the formal Lorentz pole. Increasing the resonance modulation factor from $M_{\mathrm r}=2$ to $M_{\mathrm r}=3$ substantially enhances both the total reflected power and the contributions from the two post-switch polariton branches.

\subsection{Comparing Different Modulation Techniques}
\label{comparingDMT}

The preceding results establish resonance-frequency modulation as a route to strong time reflection at a finite red detuning from the initial material resonance. We now compare this approach with the more conventional modulation of the plasma frequency while keeping the resonance frequency fixed. Plasma-frequency tuning underlies several ultrafast optical effects in transparent conducting oxides, including time refraction near the epsilon-near-zero (ENZ) regime in indium tin oxide and all-optical switching in CdO~\cite{zhou2020broadband, segal2026before}.

We characterize the two modulation protocols by $M_{\mathrm r}=\omega_{\mathrm{r},+}/\omega_{\mathrm{r},-}$ and $M_{\mathrm p}=\omega_{\mathrm{p},-}/\omega_{\mathrm{p},+}$. The reversed ratio in $M_{\mathrm p}$ is intentional: within the adopted Lorentz model, both an increase of $\omega_{\mathrm r}$ and a decrease of $\omega_{\mathrm p}$ supply energy to the inherited polarization state. Whereas increasing $\omega_{\mathrm r}$ raises only the polarization-potential energy, decreasing $\omega_{\mathrm p}$ increases both the kinetic and potential contributions, which share the prefactor $1/\omega_{\mathrm p}^{2}$ in Eq.~\eqref{eq:supp_total_energy_density}. Since $P$ and $\dot P$ remain continuous at the switch, both contributions are multiplied by $M_{\mathrm p}^{2}$. In the mechanical analogy, reducing $\omega_{\mathrm p}$ increases the effective mass in the kinetic energy expression and the restoring stiffness of the spring in the potential energy expression by the same factor, leaving the resonance frequency unchanged. Maintaining the instantaneous displacement and velocity, therefore, requires additional energy from the modulation mechanism.

To examine substantially smaller modulation than the resonance shifts considered above, here we set $M_{\mathrm r}=M_{\mathrm p}=1.12$, corresponding to a $12\%$ resonance-frequency increase or a $10.7\%$ plasma-frequency decrease. This plasma-frequency modulation value corresponds to realistic material modulation properties of CdO~\cite{segal2026before}.
We also vary the dimensionless oscillator-strength parameter $N=\omega_{\mathrm{p},-}/\omega_{\mathrm{r},-}$. All results below assume an abrupt, spatially uniform switch in a lossless Lorentz medium with $\varepsilon_\infty=1$ and an incident wave on the lower-polariton branch. We compare the two switching protocols for the same initial medium and the same multiplicative parameter contrast. It should be noted that this does not imply equal modulation work or equal pump-energy requirements.

\begin{figure}[t]
    \centering
    \includegraphics[width=0.95\linewidth]{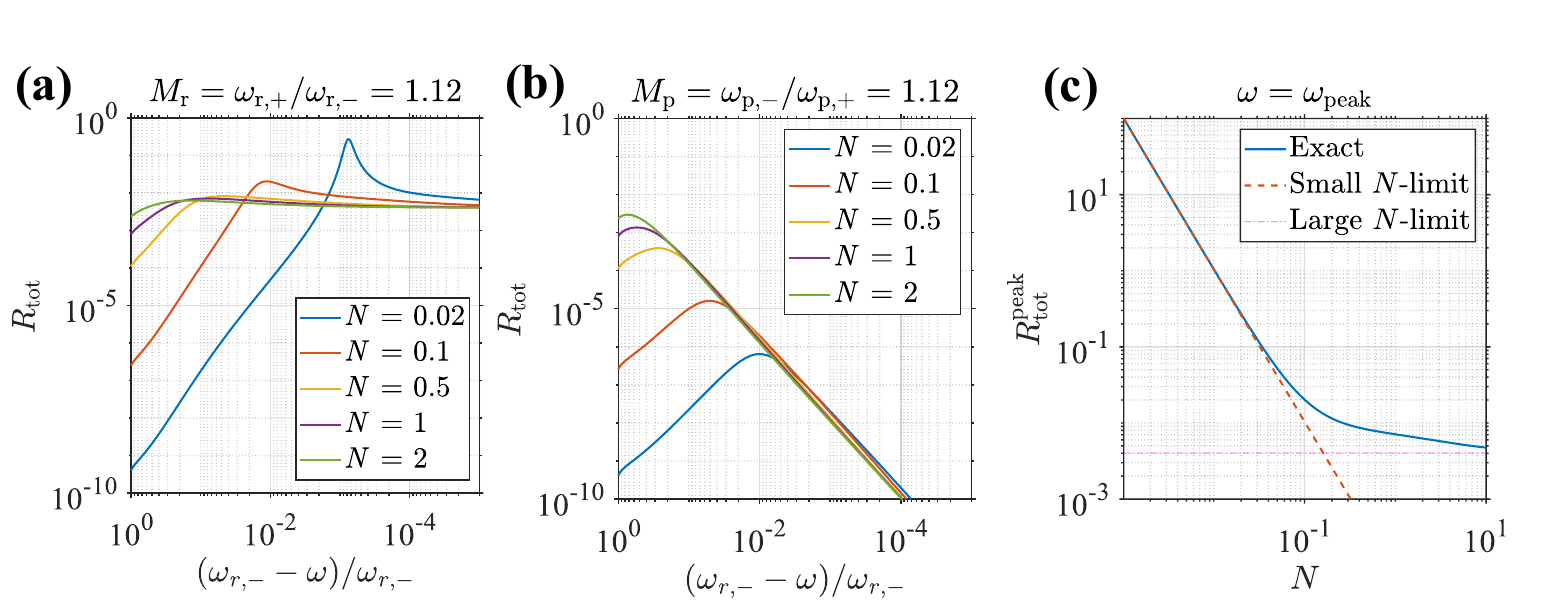}
    \caption{Comparison of resonance- and plasma-frequency modulation in a lossless Lorentz medium with $\varepsilon_\infty=1$. (a) Total reflected-power ratio $R_{\mathrm{tot}}=R_{\mathrm L}+R_{\mathrm U}$ for a resonance-frequency upshift with $M_{\mathrm r}=1.12$ at fixed plasma frequency. (b) Corresponding results for a plasma-frequency downshift with $M_{\mathrm p}=1.12$ at fixed resonance frequency. Both panels show the dependence on the normalized incident-frequency detuning for different initial oscillator-strength parameters $N=\omega_{\mathrm{p},-}/\omega_{\mathrm{r},-}$. (c) Peak reflected-power ratio at $\omega=\omega_{\rm peak}$ versus $N$ for resonance-frequency modulation with $M_{\mathrm r}=1.12$. The solid blue curve gives the exact result, the dashed orange curve shows the small-$N$ asymptote, and the horizontal pink line denotes the large-$N$ limit.}
    \label{fig:fig5}
\end{figure}

Figure~\ref{fig:fig5}(a) shows the total reflected-power ratio $R_{\mathrm{tot}}=R_{\mathrm L}+R_{\mathrm U}=\sum_{j=\mathrm L,\mathrm U}|r_j|^2\omega/\Omega_j$, normalized to the incident Poynting flux, for an abrupt resonance-frequency modulation $\omega_{\mathrm r}(t)=\omega_{\mathrm{r},-}+(\omega_{\mathrm{r},+}-\omega_{\mathrm{r},-})U(t)$ at a fixed plasma frequency$\omega_{\mathrm p}$. For every fixed $N>0$, the limit $\omega\to\omega_{\mathrm{r},-}^{-}$ gives $R_{\mathrm{tot}}\to M_{\mathrm r}(M_{\mathrm r}-1)^2/4=4.032\times10^{-3}$ (see derivation in Supplementary Section~3). The strongest time-reflection instead occurs at a finite red detuning, with an increasingly pronounced and narrow maximum as $N$ decreases.
In the weak-coupling regime, this maximum is associated with the avoided crossing of the post-switch polariton branches at a conserved wave number $ck\approx\omega_{\mathrm{r},+}$. Here, the electromagnetic and material excitations are strongly mixed, and the denominator of Eq.~\eqref{eq:modal_reflection_coefficients},  $\Omega_{\mathrm U}^{2}-\Omega_{\mathrm L}^{2}$, reaches its minimum value. The nearly resonant initial material excitation therefore projects onto two closely spaced post-switch polariton branches, each with a substantial electromagnetic fraction of its total energy.
Matching this inherited excitation requires large modal amplitudes, allowing both the initially stored energy and the energy supplied by the modulation to feed strong time-reflected waves. The incident frequency corresponding to this avoided crossing regime in the post-switch medium can be found from (see Supplementary Section~3)
\begin{equation}
\frac{\omega_{\mathrm{r},-}-\omega_{\mathrm{peak}}}
     {\omega_{\mathrm{r},-}}
\approx
\frac{N^2}{2(M_{\mathrm r}^{2}-1)}\,,
\label{eq:comparison_peak_asymptotics1}
\end{equation}
and the maximum time-reflection power ratio reaches there the value of
\begin{equation}
R_{\mathrm{tot}}^{\mathrm{peak}}
\approx
\frac{(M_{\mathrm r}^{2}-1)^2(M_{\mathrm r}-1)^2}
     {8M_{\mathrm r}N^2}\,.
\label{eq:comparison_peak_asymptotics2}
\end{equation} 
Thus, reducing $N$ while tuning the incident frequency to the corresponding peak can lead to substantial time-reflection enhancements even for modest modulation strengths. This may seem counterintuitive: for otherwise identical resonators, a smaller $N$ means a lower number density, $n_{\rm osc}\propto N^2$, suggesting less stored polarization energy. This expectation is correct only at fixed frequency detuning. Here, however, the optimal frequency $\omega_{\rm peak}$ moves progressively closer to the initial resonance as $N$ decreases, as shown in Fig.~\ref{fig:fig5}(a) and Eq.~\eqref{eq:comparison_peak_asymptotics1}. In the lossless, small-$N$ limit, this closer tuning increases the dipole-moment amplitude of each resonator as $p\propto N^{-2}$ for a fixed incident electric-field amplitude. Consequently, both the stored polarization energy density and the energy density supplied by the modulation for a fixed upward resonance shift scale as $n_{\rm osc}p^2\propto N^{-2}$, as seen from Eq.~\eqref{eq:phase_averaged_modulation_work}. The increased energy per resonator, therefore, outweighs the reduction in their number, providing a larger energy reservoir that can feed the time-reflected waves.

Plasma-frequency modulation produces a qualitatively different time-reflection response, as seen in Fig.~\ref{fig:fig5}(b). In this case, $R_{\mathrm{tot}}\to0$ as $\omega\to\omega_{\mathrm{r},-}$: approaching the unchanged material resonance suppresses the reflected power. The strongest limiting response is instead reached for $\omega\ll\omega_{\mathrm r}\ll\omega_{\mathrm{p},+}$, where both media are effectively dispersionless over the signal-frequency range. In this limit, $R_{\mathrm{tot}} \approx M_{\mathrm p}(M_{\mathrm p}-1)^2/4=4.032\times10^{-3}$ (see Supplementary Section~3). Although the finite-$N$ spectra can exhibit maxima in time-reflection power ratio, they do not display the resonant enhancement found for resonance-frequency modulation. Thus, for the plasma-frequency modulation considered here, operating in the dispersive regime weakens time reflection relative to the low-frequency, effectively dispersionless benchmark described above.

The comparison in Figs.~\ref{fig:fig5}(a) and (b) is especially striking at small $N$. For example, resonance-frequency modulation at $N=0.02$ yields a peak reflected-power ratio of approximately $0.27$, about $67$ times the effectively dispersionless benchmark $R_{\mathrm{tot}}=4.032\times10^{-3}$ approached by plasma-frequency modulation in the low-frequency, strong-oscillator regime. Figure~\ref{fig:fig5}(c) summarizes the peak resonance-modulation response as a function of $N$. The exact result follows the $N^{-2}$ scaling in Eq.~\eqref{eq:comparison_peak_asymptotics2} at small $N$ and approaches $M_{\mathrm r}(M_{\mathrm r}-1)^2/4$ as $N\to\infty$. At $N=10^{-3}$, for example, $R_{\mathrm{tot}}^{\mathrm{peak}}\approx104$, exceeding the dispersionless benchmark by approximately $2.6\times10^4$. 

These results demonstrate that selecting an appropriate excitation frequency, identifying the most suitable material parameter for modulation, and optimizing other key static parameters--such as the oscillator strength in this particular study--can be as important as increasing the modulation depth. This ideal enhancement, however, comes with an important trade-off: parameter regimes producing higher and sharper reflection peaks are generally more sensitive to material loss. Increasing the damping broadens and suppresses the peak, thereby limiting the maximum attainable reflected power. In the next section, we incorporate realistic material losses and examine a possible spatially structured material platform to determine how much of the ideal enhancement can be retained.

\section{Model Optical Platform}

\subsection{Geometry and Effective-Medium Model}

The preceding analysis identifies the oscillator-strength parameter $N=\omega_{\mathrm p}/\omega_{\mathrm r}$ as a key design variable for enhancing time reflection. In homogeneous materials, this ratio is constrained by microscopic material properties. Promising platforms for ultrafast optical modulation include dielectric semiconductors, such as germanium and silicon, and transparent conducting oxides (TCOs), such as aluminum-doped zinc oxide (AZO), indium tin oxide (ITO), and doped cadmium oxide (CdO). Silicon and germanium can combine extremely low absorption with an ultrafast nonlinear response within their infrared transparency windows. In these windows, however, their dielectric response is generally far from an intrinsic material resonance, whose frequency is difficult to modulate directly. TCOs can exhibit large optical changes near their epsilon-near-zero (ENZ) regime, but their infrared response is predominantly Drude-like, corresponding to $\omega_{\mathrm r}=0$, and includes appreciable free-carrier absorption. We therefore propose a metamaterial approach based on spatial structuring~\cite{wang2025expanding}. Structural resonances provide geometric control over the resonance frequency and oscillator strength. When an effective Lorentz description is applicable, this enables the ratio $N_{\mathrm{eff}}=\omega_{\mathrm{p},\mathrm{eff}}/\omega_{\mathrm{r},\mathrm{eff}}$ to be tailored below unity, even when the constituent material has Drude-type dispersion and only its plasma frequency can be modulated.

Two metamaterial routes based on structuring are particularly relevant. In the first approach, arrays of low-loss dielectric resonators can exploit Mie-type resonances or quasi-bound states in the continuum (qBICs), whose frequencies can be tuned by changing the constituent permittivity~\cite{wang2025expanding,garg2026photonic}. Quantitative analysis of these modes is challenging and generally requires a full-wave treatment that accounts for retardation, radiation, and coupling between resonators. In the second approach, conducting-oxide inclusions can support localized surface-plasmon resonances, arising from confined free-carrier oscillations. These resonances persist in the quasistatic limit and, when the inclusion dimensions and lattice period satisfy the homogenization conditions, admit an effective-medium description. Here, we adopt this quasistatic plasmonic route to obtain a simple and analytically tractable model connecting the constituent material properties and geometry to the \textit{effective} Lorentz response and the resulting temporal scattering.

\begin{figure}[tb]
    \centering
    \includegraphics[width=0.97\linewidth]{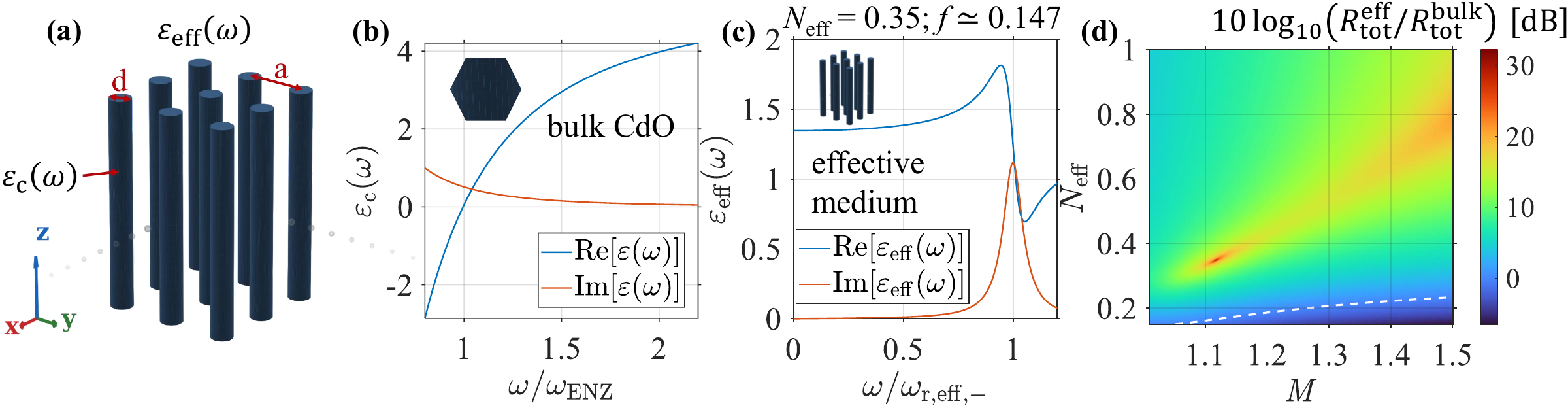}
    \caption{Model optical platform and comparison with homogeneous CdO.
(a) Square array of long CdO cylinders with diameter $d$, period $a$, and axes parallel to $z$. The effective response is considered for electric fields perpendicular to the cylinder axes.
(b) Real and imaginary parts of the pre-switch bulk CdO permittivity versus frequency normalized to the lossless ENZ reference $\omega_{\mathrm{ENZ}}=\omega_{\mathrm{p,c},-}/\sqrt{\varepsilon_\infty}$.
(c) Corresponding transverse effective permittivity for $N_{\mathrm{eff}}=0.35$ and $f=0.147$, showing the localized surface-plasmon resonance.  
(d) Modal time-reflection power enhancement versus $M$ and $N_{\mathrm{eff}}$. The cylinder array is evaluated at $\omega/\omega_{\mathrm{r,eff},-}=0.996$, while the bulk CdO material at $\omega=\omega_{\mathrm{ENZ}}$. The white dashed contour denotes the $0$~dB enhancement region.  }
    \label{fig:model}
\end{figure}

We consider a square array of identical circular cylinders in free space, as shown in Fig.~\ref{fig:model}(a). The cylinders have diameter $d$, lattice period $a$, and filling fraction $f=\pi d^2/(4a^2)$. Their axes are parallel to $z$, and they are assumed sufficiently long that end effects are negligible in the illuminated region. We consider propagation along $x$ with the electric field polarized along $y$, perpendicular to the cylinder axes. The effective permittivity derived below, therefore, describes the transverse response of the array, not its axial response. A similar theory can also be easily developed for one-dimensional Drude--dielectric multilayers or three-dimensional arrays of Drude spheres. We focus on cylinders because they provide a basic analytical model and a relatively basic geometry for nanofabrication and quasi-uniform temporal modulation. As an alternative to transparent conducting oxides, such cylinder arrays could also be realized using plasma filaments generated in air by femtosecond laser pulses, enabling subcycle temporal modulation at terahertz frequencies~\cite{gao2013femtosecond,huang2026bidirectional}.

The cylinder material in our theoretical model is described by the Drude permittivity
\begin{equation}
\varepsilon_{\mathrm c}(\omega)
=
\varepsilon_\infty
-
\frac{\omega_{\mathrm{p,c}}^{2}}
     {\omega^{2}+i\gamma\omega}\,,
\label{eq:cylinder_response}
\end{equation}
where $\varepsilon_\infty$ is the real, nondispersive background permittivity of the cylinders, $\omega_{\mathrm{p,c}}$ is their plasma frequency, and $\gamma$ is the damping rate. This is the special case of Eq.~\eqref{eq:Lorentz_susceptibility} with $\omega_{\mathrm r}=0$. For simplicity, we assume that the surrounding host is free space.

As mentioned above, we adopt a quasistatic model and retain only the transverse electric-dipole response of each cylinder. Neglecting retardation within a cylinder and across a unit cell requires $|k_{\mathrm c}|d\ll1$, $k_0a\ll1$, and $|k_{\mathrm{eff}}|a\ll1$ ~\cite[p.~151]{sihvola1999electromagnetic}, where $k_0=\omega/c$, $k_{\mathrm c}=k_0\sqrt{\varepsilon_{\mathrm c}}$, and $k_{\mathrm{eff}}=k_0\sqrt{\varepsilon_{\mathrm{eff}}}$. The first condition requires the cylinder diameter to be much smaller than both the internal wavelength and the field-penetration depth, including the skin depth when the internal field is evanescent. These conditions must hold at the incident frequency and at the frequencies of the appreciably excited post-switch modes. 
Nevertheless, these restrictions concern only the chosen simple analytical model, not the underlying energy-storage mechanism. Resonance-enhanced time reflection can likewise be explored in low-loss dielectric resonators beyond the quasistatic regime (e.g., supporting Mie resonances), provided that temporal switching couples the initially stored resonant energy into backward-propagating modes.

Although the constituent Drude response has no finite restoring frequency, a transverse displacement of its free carriers creates surface charges on each cylinder. Their depolarizing field provides a restoring force and produces a finite-frequency structural resonance. Within the Maxwell--Garnett approximation, the resulting transverse effective permittivity has the Lorentz form~\cite{sihvola1999electromagnetic}
\begin{equation}
\varepsilon_{\mathrm{eff}}(\omega)
=
\varepsilon_{\infty,\mathrm{eff}}
+
\frac{\omega_{\mathrm{p,eff}}^{2}}
     {\omega_{\mathrm{r,eff}}^{2}-\omega^{2}-i\gamma\omega}\,,
\label{eq:effective_Lorentz}
\end{equation}
with
\begin{equation}
\begin{aligned}
\varepsilon_{\infty,\mathrm{eff}}
&=
\frac{(1-f)+(1+f)\varepsilon_\infty}
     {(1+f)+(1-f)\varepsilon_\infty}\,,
\\[4pt]
\omega_{\mathrm{p,eff}}
&=
\frac{2\sqrt{f}}
     {(1+f)+(1-f)\varepsilon_\infty}\,
\omega_{\mathrm{p,c}}\,,
\\[4pt]
\omega_{\mathrm{r,eff}}
&=
\sqrt{\frac{1-f}
     {(1+f)+(1-f)\varepsilon_\infty}}\,
\omega_{\mathrm{p,c}}\,.
\end{aligned}
\label{eq:effectiveParameters}
\end{equation}
The effective damping rate remains $\gamma$ within this quasistatic approximation.
The corresponding oscillator-strength parameter is
\begin{equation}
N_{\mathrm{eff}}
=
\frac{\omega_{\mathrm{p,eff}}}{\omega_{\mathrm{r,eff}}}
=
2\sqrt{\frac{f}
{(1-f)\bigl[(1+f)+(1-f)\varepsilon_\infty\bigr]}}\,.
\label{eq:N_eff}
\end{equation}
Reducing the filling fraction therefore decreases $N_{\mathrm{eff}}$ without changing the constituent material. In the dilute limit, $N_{\mathrm{eff}}\approx2\sqrt{f/(1+\varepsilon_\infty)}$, explicitly demonstrating how geometry can provide access to $N_{\mathrm{eff}}\ll1$. Similarly to the explanation in Section~\ref{comparingDMT}, although dilution reduces the amount of resonant material, retuning the incident frequency to the corresponding time-reflection peak drives the cylinders increasingly close to resonance. The resulting stronger resonant excitation can overcompensate for the reduced filling fraction, increasing the stored polarization energy at a fixed incident-field amplitude.

For the optical material example considered below, we use CdO cylinders with $\varepsilon_\infty=5.3$, $\omega_{\mathrm{p,c},-}=2\pi\times157.6$~THz, and $\gamma=0.0424\omega_{\mathrm{p,c},-}$ \cite{segal2026before}. 
The frequency dispersion of CdO is shown in Fig.~\ref{fig:model}(b).
During each temporal switching event, we keep the geometry, $f$, $\varepsilon_\infty$, and $\gamma$ fixed and modulate only the cylinder plasma frequency. We describe the abrupt, spatially uniform change by $\omega_{\mathrm{p,c}}(t)=\omega_{\mathrm{p,c},-}[1+(M-1)U(t)]$, where $M=\omega_{\mathrm{p,c},+}/\omega_{\mathrm{p,c},-}$. Equation~\eqref{eq:effectiveParameters} then gives $\omega_{\mathrm{r,eff},+}=M\omega_{\mathrm{r,eff},-}$ and $\omega_{\mathrm{p,eff},+}=M\omega_{\mathrm{p,eff},-}$, while $\varepsilon_{\infty,\mathrm{eff}}$ and $N_{\mathrm{eff}}$ remain unchanged. Thus, a single material-control parameter simultaneously changes both effective Lorentz frequencies. This protocol differs from the resonance-frequency modulation at fixed plasma frequency analyzed in Sec.~\ref{energyexch}. In the notation of Sec.~\ref{comparingDMT}, it corresponds to $M_{\mathrm r}=M$ and $M_{\mathrm p}=M^{-1}$. The temporal dynamics are described by Eq.~\eqref{eq:Lorentz_time_domain} with these effective parameters, retaining the continuity of $E$, $H$, $P$, and $\dot P$ under the adopted switching prescription. The derivation of the time-reflected power coefficient in this case can be found in Supplementary Section~4.

\subsection{Comparison with the Bulk Homogeneous Material}
\label{subsec:bulk_comparison}

We next compare the time-reflection strength of the CdO cylinder array with that of homogeneous bulk CdO under the same relative change in the constituent plasma frequency. This comparison quantifies the benefit of introducing an effective Lorentz resonance through spatial structuring.

Figure~\ref{fig:model}(c) shows the pre-switch effective permittivity for an array with $N_{\mathrm{eff}}=0.35$. For the chosen CdO background permittivity, Eq.~\eqref{eq:N_eff} uniquely determines the filling fraction as $f=0.147$, corresponding to $d/a=2\sqrt{f/\pi} = 0.433$. A representative geometry is therefore $d=26$~nm and $a=60$~nm, providing a plausible target for nanofabrication. For the operating regime considered below, the largest values of $|k_{\mathrm c}|d$, $|k_0|a$, and $|k_{\mathrm{eff}}|a$ over the incident and four post-switch modes are approximately $0.074$, $0.089$, and $0.100$, respectively (much less than unity).  The effective response exhibits a localized surface-plasmon resonance, with a pronounced absorption peak and rapid variation of the real permittivity near $\omega_{\mathrm{r,eff},-}$ (see Fig.~\ref{fig:model}(c)). This finite-frequency resonance, absent from the homogeneous Drude response in Fig.~\ref{fig:model}(b), enables resonant polarization-energy storage before the temporal switch.

To quantify the effect of structuring, Fig.~\ref{fig:model}(d) compares the array of CdO cylinders and homogeneous bulk CdO material under the same constituent plasma-frequency change $M$. Both time-reflected power coefficients are calculated using the formulation explained in Supplementary Section~4.  The array is evaluated at $\omega=0.996\omega_{\mathrm{r,eff},-}$,
whereas the bulk reference is evaluated at its ENZ region, i.e., at $\omega= \omega_{\mathrm{p,c},-}/\sqrt{\varepsilon_\infty}$. Thus, the comparison uses the respective near-resonance and near-ENZ operating frequencies, with each reflected-power coefficient normalized to its own incident power flux.

As shown in Fig.~\ref{fig:model}(d), the time-reflected power enhancement reaches approximately $30$~dB in the narrow red region. For the representative array in panel~(c), with $f=0.147$, we consider a modulation strength of $M=1.12$, motivated by the experimentally demonstrated ultrafast plasma-frequency tuning of CdO~\cite{segal2026before}. The resulting total time-reflected power coefficients are $R_{\mathrm{tot}}^{\mathrm{eff}}=3.98$ and $R_{\mathrm{tot}}^{\mathrm{bulk}}= 2.38\times10^{-3}$, corresponding to an enhancement of approximately $32.2$~dB.
The sharp peak in time-reflection coefficient for the cylinder array arises from proximity to an exceptional point of the post-switch temporal eigenproblem evaluated at the conserved complex wavevector, where two complex eigenfrequencies and their eigenvectors coalesce.  Related coalescence and cancellation of temporal-scattering coefficients were discussed in~\cite{solis2021time}.
The enhancement in Fig.~\ref{fig:model}(d) refers to the powers of the two time-reflected modes considered separately. Near the sharp maximum, their frequencies become close and their electric fields are nearly opposite in phase, causing substantial cancellation while the reflected pulses overlap. Their different group velocities can enable separation during propagation, as illustrated for the lossless case in Fig.~\ref{fig:FDTD2}, but absorption may attenuate the pulses before they become distinguishable in CdO metamaterial. The quantitative analysis is provided in Supplementary Section~4. This introduces a trade-off: approaching the enhancement maximum increases the individual modal contributions but makes them harder to resolve experimentally. Choosing an operating point farther from the maximum can alleviate this difficulty while retaining a smaller enhancement over homogeneous CdO.

A possible experimental route is to use a finite array slab and arrange the temporal switch while the incident pulse is close to a spatial interface with air. For oblique incidence in the plane perpendicular to the cylinder axes, the generated time-reflected components return toward this interface and can be extracted after a short propagation distance through the lossy array. Subsequent propagation in air avoids further material absorption. Conservation of frequency and tangential wavevector at the stationary interface also permits slightly different real-frequency components to emerge at slightly different angles, suggesting angularly resolved detection. This geometry could reduce absorption losses and assist discrimination of the two reflected modes. Nevertheless, the angular separation must exceed their angular widths. Quantifying the experimentally accessible enhancement therefore requires a finite-pulse calculation for the complete slab, including interface transmission, spectral and angular overlap, and the influence of the boundaries on the modes. Figure~\ref{fig:model}(d) identifies promising parameters for such an investigation, while collection of the full modal enhancement remains a question of excitation and outcoupling design.

\section{Conclusion}

In this work, we have established a resonance-assisted approach to enhancing time reflection at a single temporal interface. Analytical calculations, supported by finite-difference time-domain simulations, show that resonance-frequency modulation can produce substantially stronger reflection than plasma-frequency modulation, with enhancements of several orders of magnitude possible both in the lossless and lossy regimes. Our energy-balance analysis and mechanical oscillator analogy explain this effect through the redistribution of resonantly stored polarization energy and the work exchanged with the modulation mechanism. The post-switch polariton structure governs how this energy from the modulation mechanism feeds the reflected waves, making oscillator strength and incident detuning key design parameters, while material losses limit the attainable enhancement. 
Finally, an effective-medium model of CdO cylinder arrays predicts an enhancement of the sum of the time-reflected modal powers by several orders of magnitude relative to homogeneous CdO under the same constituent plasma-frequency modulation, even with realistic material losses. This enhancement arises from the controllable Lorentz resonance introduced into the Drude response through spatial structuring.

The resonance-assisted mechanism suggests several routes toward experimental implementation. At microwave frequencies, time-modulated transmission lines, lumped-element circuits, and tunable resonant metasurfaces provide accessible platforms for testing rapidly switched dispersive responses~\cite{moussa2023observation,wang2023metasurface,Li2025NonFoster,Li2026Multimode}. At terahertz and optical frequencies, ultrafast modulation can exploit free-carrier injection in semiconductors~\cite{rawat2026generation}, laser-induced plasma formation~\cite{gao2013femtosecond,huang2026bidirectional}, and changes in the free-carrier response of transparent conducting oxides~\cite{zhou2020broadband,segal2026before}. Combining these mechanisms with subwavelength resonators, metasurfaces, and photonic-crystal slabs could enable rapid control of engineered electromagnetic resonances, with even modest material changes producing shifts comparable to or exceeding the resonance linewidth. Long-lived resonances in low-loss structures, including quasi-bound states in the continuum, may further increase the energy stored before switching~\cite{fan2002analysis,yanik2004stopping,garg2026photonic}, provided that efficient excitation and coupling to the outgoing waves are ensured. These possibilities offer promising routes toward the experimental realization of strong resonance-enhanced time reflection at terahertz and optical frequencies.

\section*{Acknowledgments}
V.A. acknowledges the Finnish Foundation for Technology Promotion, Horizon Europe Marie Skłodowska-Curie Actions (MSCA) program (no. 101226375), and Research Council of Finland (no.  365679 and 371367). P.G. and C.R. are part of the Max Planck School of Photonics, supported by the Bundesministerium für Bildung und Forschung, the Max Planck Society, and the Fraunhofer Society. P.G. and C.R. acknowledge support by the German Research Foundation within the SFB 1173 (project ID no. 258734477). P.G. acknowledges support from the Karlsruhe School of Optics and Photonics (KSOP). X.W. and Z.L. acknowledge financial support from the Fundamental Research Funds for the Central Universities, China (Project No. 3072026RL2501), and the National Natural Science Foundation of China (Grant Nos. 62541116 and 12674385). M.H.M. acknowledges the support from the Walter Ahlstrom foundation in Finland.

\section{Appendix: Methods}
\label{sec:methods}

\paragraph{Analytical temporal-scattering calculation.}
For a prescribed incident frequency \(\omega\), the conserved wave number is determined from the dispersion relation of the initial Lorentz medium,
\begin{equation}
k^2
=
\frac{\omega^2}{c^2}
\varepsilon_{\mathrm r,-}(\omega)\,.
\label{eq:initial_lorentz_dispersion}
\end{equation}
Here, we assume that the incident wave is represented by a single (lower polaritonic) mode at frequency $\omega$. 
Because the temporal modulation is spatially uniform, this wave number remains unchanged across the temporal interface. The post-switch eigenfrequencies are then obtained by solving
\begin{equation}
k^2
=
\frac{\Omega_\nu^2}{c^2}
\varepsilon_{\mathrm r,+}(\Omega_\nu)
\label{eq:post_switch_lorentz_dispersion}
\end{equation}
at the same \(k\). For the Lorentz-to-vacuum switch, the post-switch spectrum contains the pair \(\Omega=\pm c k\), corresponding to one time-refracted and one time-reflected wave. Their electric-field amplitudes are obtained by preserving the total electric and magnetic fields across the interface, leading to Eq.~\eqref{eq:vacuum} (see Supplementary Section~2).

For a finite Lorentz-to-Lorentz resonance shift, Eq.~\eqref{eq:post_switch_lorentz_dispersion} yields lower- and upper-polariton solutions together with their negative-frequency counterparts. The inherited field--matter state is expanded over these $\nu=1, 2, 3, 4$ post-switch modes. Because \(\varepsilon_\infty\) and \(\omega_{\mathrm p}\) remain unchanged in this switching scenario, the modal amplitudes are determined from the complete temporal-interface conditions in Eq.~\eqref{eq:complete_temporal_boundary_conditions}, namely the continuity of \(D\), \(B\), \(P\), and \(\dot{P}\). The coefficients \(T_{\rm L}\), \(T_{\rm U}\), \(R_{\rm L}\), and \(R_{\rm U}\) are defined as the powers of the resulting modes normalized to the incident power. A detailed derivation of the modal electric-field and power coefficients is provided in Supplementary Section~2.

\paragraph{ADE--FDTD implementation.}
We perform the time-domain simulations using an in-house one-dimensional FDTD solver. The dispersive material response is incorporated through the standard auxiliary-differential-equation (ADE) method. 
For the finite Lorentz-to-Lorentz transitions, \(\omega_{\mathrm r}(t)\) is changed from \(\omega_{\mathrm r,-}\) to \(\omega_{\mathrm r,+}\), while \(\varepsilon_\infty\), \(\omega_{\mathrm p}\), and the instantaneous ADE state \((P,J)\) are retained across the switch. This implementation preserves the polarization and polarization current and therefore realizes the temporal-interface conditions used in the analytical model. For the limiting Lorentz-to-vacuum benchmark, the ADE polarization branch is removed at the temporal interface, and the post-switch domain is updated as vacuum, consistently with the field-preserving switching prescription used to obtain Eq.~\eqref{eq:vacuum}. In both cases, the modulation is applied uniformly throughout the computational domain over a transition interval much shorter than the incident-wave period, thereby approximating an abrupt temporal interface and preserving the wave number.

A narrowband, sinusoidally modulated Gaussian pulse with carrier frequency $\omega$ is launched into the initially time-invariant Lorentz medium. The source waveform is defined as
$E_{\mathrm{src}}(t)
=
E_0
\exp\!\left[-\frac{(t-T_{\mathrm c})^2}{2T_{\mathrm d}^2}\right]
\sin(\omega t),$
where the temporal full width at half maximum is $\mathrm{FWHM}=15~\mathrm{ns}$, $T_{\mathrm d}=\mathrm{FWHM}/[2\sqrt{2\ln(2)}]$, and the pulse center is set to $T_{\mathrm c}=20~\mathrm{ns}$. The temporal switch is applied after the pulse has entered the medium at 300~ns. Perfectly matched layers truncate both ends of the computational domain and are terminated by perfect electric conductor (PEC) boundaries to suppress reflections from the spatial boundaries. The electric field \(E(x,t)\) is recorded throughout the simulation to obtain the spatiotemporal maps shown in Figs.~\ref{fig:FDTD1} and \ref{fig:FDTD2}. For these maps, the field is normalized by the peak incident electric-field amplitude immediately before switching.

\paragraph{Extraction of temporal-scattering amplitudes.}
After the temporal interface, the generated wave packets propagate in different directions and at different group velocities. They are allowed to separate spatially before their amplitudes are evaluated. For each isolated channel, the FDTD amplitude is defined by
\begin{equation}
\left|r_j\right|_{\mathrm{FDTD}}
=
\frac{
\displaystyle\max_{\mathcal{W}_{r_j}}|E(x,t)|
}{
\displaystyle\max_{\mathcal{W}_{\mathrm{inc}}}|E(x,t)|
}\, ,
\qquad
\left|t_j\right|_{\mathrm{FDTD}}
=
\frac{
\displaystyle\max_{\mathcal{W}_{t_j}}|E(x,t)|
}{
\displaystyle\max_{\mathcal{W}_{\mathrm{inc}}}|E(x,t)|
}\, ,
\label{eq:FDTD_peak_coefficients}
\end{equation}
where \(\mathcal{W}_{r_j}\) and \(\mathcal{W}_{t_j}\) are nonoverlapping spatial domains containing the corresponding reflected and transmitted wave packets, and \(\mathcal{W}_{\mathrm{inc}}\) contains the incident pulse immediately before the switch. Here, \(j={\rm U,L}\) defines the mode subscript after the temporal jump.

The carrier frequencies and relative phases of the separated wave packets are obtained by Fourier transforming the gated electric-field data. Figures~\ref{fig:FDTD2}(c) and \ref{fig:FDTD2}(d) display the magnitudes of the four modal coefficients. Because the incident pulse is narrowband, the extracted peak-amplitude ratios can be directly compared with the monochromatic analytical coefficients evaluated at the pulse carrier frequency.

\bibliographystyle{ieeetr}
\bibliography{references}

\end{document}